\documentclass[prL,a4paper,superscriptaddress,twocolumn,amsmath,amssymb]{revtex4}

\usepackage{graphicx}
\usepackage{bm}
\graphicspath{{./Figs/}}
\usepackage{dsfont}
\usepackage{epstopdf}
\usepackage{hyperref}
\usepackage[squaren]{SIunits}
\usepackage{color}
\newcommand {\Fig}[1] {Figure~\ref{#1}}

\newcommand{\beq}{\begin{equation}}
\newcommand{\eeq}{\end{equation}}

\newcommand{\sep}{Se$^{+}$}

\newcommand{\beqa}{\begin{eqnarray}}
\newcommand{\eeqa}{\end{eqnarray}}

\newcommand{\ttwo}{$T_2$}

\newcommand{\tone}{$T_1$}

\begin{document}

\title{Spin relaxation and donor-acceptor recombination of Se$^+$ in 28-silicon}

\author{Roberto Lo Nardo}
\affiliation{London Centre for Nanotechnology, University College London, London WC1H 0AH, UK} 
\affiliation{Dept.\ of Materials, Oxford University, Oxford OX1 3PH, UK} 

\author{Gary Wolfowicz}
\affiliation{London Centre for Nanotechnology, University College London, London WC1H 0AH, UK} 
\affiliation{Dept.\ of Materials, Oxford University, Oxford OX1 3PH, UK} 

\author{Stephanie Simmons}
\affiliation{Dept.\ of Materials, Oxford University, Oxford OX1 3PH, UK} 

\author{Alexei M. Tyryshkin}
\affiliation{Dept.\ of Electrical Engineering, Princeton University, Princeton, New Jersey 08544, USA}

\author{Helge Riemann}
\affiliation{Institute for Crystal Growth, Max-Born Strasse 2, D-12489 Berlin, Germany} 

\author{Nikolai V. Abrosimov}
\affiliation{Institute for Crystal Growth, Max-Born Strasse 2, D-12489 Berlin, Germany} 

\author{Peter Becker}
\affiliation{Physikalisch-Technische,  D-38116 Braunschweig, Germany} 

\author{Hans-Joachim Pohl}
\affiliation{Vitcon Projectconsult GmbH,  07745 Jena, Germany} 

\author{Michael Steger}
\affiliation{Dept.\ of Physics, Simon Fraser University, Burnaby, British Columbia V5A 1S6, Canada}

\author{Stephen A. Lyon}
\affiliation{Dept.\ of Electrical Engineering, Princeton University, Princeton, New Jersey 08544, USA}

\author{Mike L. W. Thewalt}
\affiliation{Dept.\ of Physics, Simon Fraser University, Burnaby, British Columbia V5A 1S6, Canada}

\author{John~J.~L.~Morton}
\email{jjl.morton@ucl.ac.uk}
\affiliation{London Centre for Nanotechnology, University College London, London WC1H 0AH, UK} 
\affiliation{Dept.\ of Electronic \& Electrical Engineering, University College London, London WC1E 7JE, UK}

\begin{abstract}
Selenium impurities in silicon are deep double donors and their optical and electronic properties have been recently investigated due to their application for infrared detection. However, a singly-ionised selenium donor (Se$^{+}$) possesses an electron spin which makes it a potential candidate as a silicon-based spin qubit, with significant potential advantages compared to the more commonly studied group V donors. Here we study the electron spin relaxation (\tone) and coherence (\ttwo) times of Se$^{+}$ in isotopically purified 28-silicon, and find them to be up to two orders of magnitude longer than shallow group V donors at temperatures above $\sim \unit{15}{K}$. We further study the dynamics of donor-acceptor recombination between selenium and boron, demonstrating that it is possible to control the donor charge state through optical excitation of neutral Se$^0$.
\end{abstract}

\maketitle


\section{Introduction}
The electron and nuclear spins of Group V donors in silicon, such as phosphorus, have been actively studied as potential quantum bits~\cite{Morton2011, Awschalom2013}. Recent advances in this field include  the manipulation and read-out of individual electron and nuclear spins by integrating donors into nanoelectronic devices \cite{Morello2010, Pla2012, Pla2013}, and the demonstration that donor spin coherence times can be as long as 3 seconds for the electron spin~\cite{Wolfowicz2013} and up to 3 hours for the nuclear spin~\cite{Saeedi2013}. In contrast to such work on shallow donors in silicon as qubits, there has been relatively little experimental attention on the chalcogens, which are deep double donors in silicon. Nevertheless, they possess several attributes of potential relevance to donor-based spin qubits --- we focus here on selenium as an example.

First, neutral selenium (Se$^0$) possesses two bound electrons which form a singlet ground state, and has a binding energy of \unit{307}{meV}~\cite{Janzen1984}. This makes it ideal for spin readout methods using spin-to-charge conversion that require that the two-electron state on the donor is well bound~\cite{Kane1998, Kane2000}. In phosphorus, for example, the two-electron state (P$^-$) has a very weak binding energy of only \unit{2}{meV}. Second, the singly-ionised state, Se$^+$, is isoelectronic with the neutral Group V donors and possesses a bound $S=1/2$ electron spin. Compared to shallow donors, Se$^+$ has a much larger thermal ionisation energy (\unit{593}{meV})~\cite{Grimmeiss1982} such that it can retain a bound electron even at room temperature. Third, the large energy separation between the donor ground state, $1s(A_1)$, and first excited valley state, $1s(T_2)$ is of order 429~meV (more than an order of magnitude greater than for the shallow Group V donors) and could be expected to result in a significantly longer electron spin relaxation time.

Here we present a study of the electron spin properties of Se$^+$ in isotopically purified 28-silicon, including timescales for  electron spin relaxation (\tone) and decoherence (\ttwo), and their mechanisms. We find that the electron spin relaxation times are as much as two orders of magnitude longer than phosphorus, for a given temperature, and that electron spin decoherence times can be reasonably expected to be as long as those measured for phosphorus (up to seconds~\cite{Tyryshkin2012}). We go on to investigate donor-acceptor (DA) recombination following above band gap illumination (\unit{1047}{nm}), by monitoring the electron spin echo intensity, using the same method previously applied to phosphorus donors~\cite{Dirksen1989}. This demonstrates an optical mechanism for placing the donors in the neutral Se$^0$ state, thus removing the hyperfine interaction between the electron and the nucleus. { \color{black} We find DA recombination is slow (minutes to hours) given the concentrations of selenium ($\sim5\times 10^{15}$~cm$^{-3}$) and boron ($\sim 5 \times 10^{13}$~cm$^{-3}$) used here \cite{Dirksen1989} and the rate at which charge equilibrium is re-established can be enhanced by selectively ionising the Se$^0$ via optical illumination at \unit{4}{\mu m}.}

\section{Materials and Methods}\label{materials}
The $^{28}$Si:Se samples used here have been previously measured by IR absorption spectroscopy~\cite{Steger2009}, and consist of $^{28}$Si doped with selenium, and partially compensated with boron in order to produce a significant concentration of Se$^+$, as described by \citet{Ludwig1965}. The starting silicon material had a composition of 99.991\% $^{28}$Si, with \unit{75}{ppm} $^{29}$Si and \unit{15}{ppm} $^{30}$Si{\color{black}, and a residual boron concentration [B] $\sim 5 \times 10^{13}$~cm$^{-3}$. This was then doped with selenium by thermal diffusion}~\cite{Janzen1984}. We studied two samples, one with natural abundance selenium, and the other made using $^{77}$Se with an isotopic enrichment of 97.1 \%. The sample resistivity was measured to be 4.4~\ohm cm which corresponds to a donor concentration of $\sim5\times 10^{15}$~cm$^{-3}$, making the samples \emph{n-type}.  ESR measurements were carried out using a Bruker Elexsys 580 spectrometer at X-band (\unit{9.7}{GHz}).

\section{Continuous wave ESR}
\begin{figure}[t]
\begin{center}
\includegraphics[width=250pt]{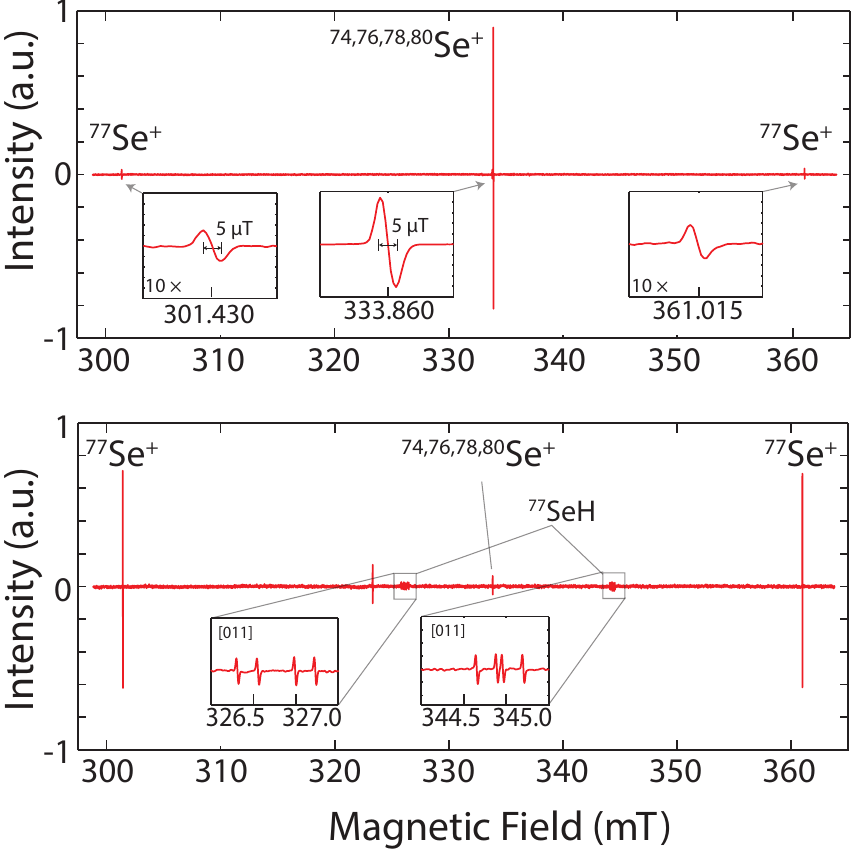}
\caption{X-band cw-ESR of Se$^+$ in $^{28}$Si at X-band for $^{28}$Si:$^{\text{nat}}$Se (upper) and $^{28}$Si:$^{77}$Se (lower). The natural abundance of $^{77}$Se is 7.5$\%$, with the remaining isotopes (92.5$\%$ abundance) possessing zero nuclear spin. The $^{77}$Se-doped sample shows predominantly the hyperfine-split lines arising from coupling to the $I=1/2$ nuclear spin of $^{77}$Se, while addition features in the spectrum correspond to residual isotopes of Se and the presence of SeH pairs. Microwave frequency = 9.38~GHz; temperature = 23~K.
 \label{cwEPR}}
\end{center}
\end{figure}

The ESR proprieties of Se$^+$ in silicon have been previously investigated by Grimmeiss \emph{et al.}~\cite{Grimmeiss1981,Janzen1984}. The electronic $g$-factor is $g=2.0057$. All  stable Se isotopes $^\text{X}$Se$^+$ (X=74, 76, 78, 80) have zero nuclear spin, apart from $^{77}$Se which has $I=1/2$ and an isotropic hyperfine coupling of $A=\unit{1.6604}{GHz}$ with the donor electron spin. The electron and nuclear spin  proprieties of Se$^+$ are described by the spin Hamiltonian (in frequency units):
\begin{equation}
{\cal H}_0=\omega_e S_z - \omega_n I_z + A \, \boldsymbol{S}\cdot\boldsymbol{I}\label{hamiltonian}
\end{equation}
where $\omega_e= g\beta B_0/\hbar$ and $\omega_n= g_n \beta_n B_0/\hbar$ are the electron and nuclear Zeeman frequencies and $B_0$ is the static magnetic field applied along the $z$ axis. Neutral Se$^0$ has a spin singlet ground state and thus gives no contribution to the ESR signal. 
Figure \ref{cwEPR} shows continuous wave ESR spectra for $^{28}$Si:$^{\text{nat}}$Se and $^{28}$Si:$^{77}$Se. The values we obtain for $A$ and $g$ confirm previously reported results \cite{Grimmeiss1981}, though the transitions have a much smaller linewidth  ($<5~\mu$T) due to silicon isotopic purification. The relative intensity of the central line (around $g=2$, corresponding to Se isotopes with zero nuclear spin) and hyperfine-split lines (corresponding to $^{77}$Se) in each sample is consistent with their expected isotopic composition. The remaining features in the $^{77}$\sep\ ESR spectrum have been characterised by angular dependent cw-ESR and ENDOR  and confirmed to be due to SeH pairs, previously measured in natural Si and referred as Si-NL60~\cite{Huy2000}. The presence of the hydrogen in this defect centre reduces the $^{77}$Se hyperfine coupling by a factor of $\sim$3 and gives it a slight anisotropy.  


\section{Electron spin relaxation}

\begin{figure}[t]
\begin{center}
\includegraphics[width=230pt]{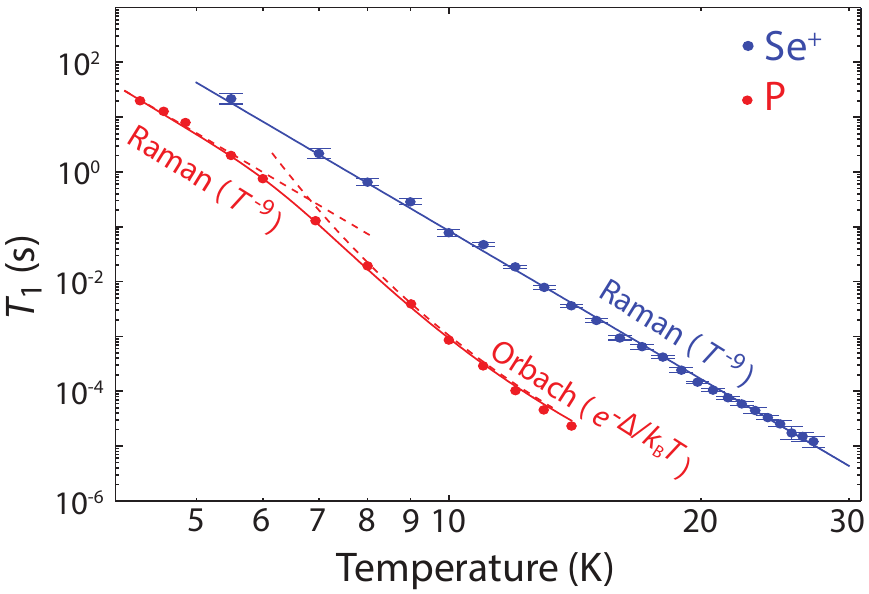}
\caption{Temperature dependence of electron spin relaxation $T_1$ in $^{28}$Si:$^{77}$Se$^+$ and Si:P at $B\sim \unit{0.34}{T}$. {\color{black} Solid lines are fits to the experimental data, made up of contributions (dashed lines) from different relaxation mechanisms.} Data for Si:P is from Ref~\cite{Castner1963}. \label{T1}}
\end{center}
\end{figure}

We studied the electron spin relaxation time ($T_1$) of Se$^+$ by pulsed ESR, using an inversion-recovery sequence ($\pi-T-\pi/2-\tau-\pi-\tau-$echo, where $T$ is varied). The results, shown in Figure \ref{T1}, show that $(1/T_1)$ is well fit by $1/T_1=CT^9$ with $C=\unit{1.2\times10^{-8}}{s^{-1}K^9}$, in the temperature range of \unit{5-35}{K}, and is independent {\color{black} of the projection of selenium nuclear spin ($m_I=0, \pm1/2$)}. A comparison with P donors in silicon shows that \tone\ is between one to two orders of magnitude longer for Se$^+$ in this temperature range.

The temperature dependence of the electron spin $T_1$ has been well studied for shallow donors \cite{Castner1963}, with different mechanisms dominating in different regimes of temperature and magnetic field. The Orbach two-phonon relaxation process~\cite{ Orbach1961, Castner1967} {\color{black} depends exponentially on} the energy gap, $\Delta = 1s(T_2)-1s(A_1)$, between the ground state and first excited state of the donor: \tone\ $\propto \exp(\Delta/k_BT)$. Therefore, though the {\color{black} Orbach process} is the dominant \tone\ mechanism for phosphorus donors at X-band and at temperatures above $\sim$8~K, the large value of $\Delta = 429$~meV for \sep\ makes the Orbach process irrelevant here. A  $T^{9}$ Raman process dominates the phosphorus donor electron spin relaxation in the range $\unit{2-6}{K}$ at X-band~\cite{Castner1967},  arising from two-phonon scattering via a continuum of excited states. It has been observed in the shallow donors (see Table~\ref{Table}) that this process has only a weak dependence on $\Delta$, {\color{black} which dictates the spin mixing through a spin-orbit coupling.}
Our results on \sep\ are consistent with this observation ---  despite the much larger value of $\Delta$, the strength of the $T^9$ Raman process is only marginally weaker than for the shallow donors. Nevertheless, the electron spin \tone\ for \sep\ in silicon remains  longer than any of the shallow donors across the full temperature range studied here, suggesting that it may also offer the longest spin coherence times.

\begin{table}
\begin{center}
\begin{tabular}{ccccc}
\hline\hline   
Donor   & Raman temp-   & $C$ (s$^{-1}$K$^{9}$) &  $\Delta$   (meV) & Reference  \\
&erature range (K) & $CT^9$&\,   & \\
\hline
P         &  2.6 - 6    &  $10\times10^{-8}$    & 10.5  &  \cite{Castner1963}  \\
As        &5 - 11  & $2\times10^{-8}  $       &  19.7    & \cite{Castner1963} \\
Bi         &5 - 26  &$ 6.6\times10^{-8}$    &   33.8  &\cite{Wolfowicz2012a} \\
Se$^+$ & 5 - 35 & $1.2\times10^{-8}   $    & 429 & this work      \\
\hline\hline
\end{tabular}
\caption{Comparison of the strengths of the $T^9$ Raman electron spin relaxation process for various shallow donors and \sep\ in silicon, including the temperature range at which this mechanisms dominates $T_1$ at X-band. Only a weak dependence is observed between the strength of the process, $C$, and the energy separation ($\Delta$) between the ground $1s(A_1)$ and first excited state ($1s(T_2)$) of the donor.\label{Table}}
\end{center}
\end{table}

\begin{figure}[t]
\begin{center}
\includegraphics[width=250pt]{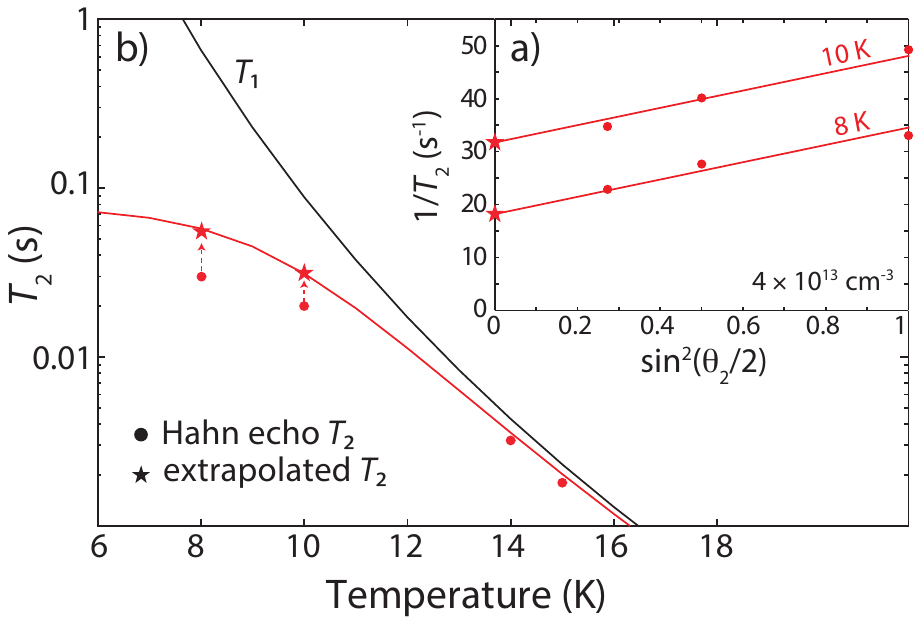}
\caption{\color{black} Measured and extrapolated $T_2$ in $^{28}$Si:$^{77}$Se$^+$ at $B\sim \unit{0.34}{T}$. a) $1/T_2$ as function of the refocusing pulse rotation angle $\theta_2$. The intercept of the linear fit gives the extrapolated $T_2$ corresponding to the suppression of the instantaneous diffusion (ID), while the slope gives a Se$^+$ concentration of \unit{4\times10^{13}}{cm^{-3}}. b) 
The value of $T_2$ is limited by $T_1$ above $\sim$\unit{12}{K}. Below this temperature, the extrapolated value of \ttwo\ (in the absence of ID) can be well described by a combination of spectral diffusion arising from \tone-induced spin flips of neighbouring \sep\ spins, combined with a temperature-independent mechanism that limits $T_2$ to about \unit{80}{ms}. \label{T2}}
\end{center}
\end{figure}

{ \color{black} Electron spin coherence times ($T_2$) of $^{77}$Se$^+$ were measured using a Hahn echo sequence ($\pi/2- \tau - \pi - \tau - \text{echo}$) and are shown in \Fig{T2}}. We find that $T_2$ is limited by spin relaxation ($T_1$) for temperatures above $\sim$\unit{12}{K}. 
Below this temperature, $T_2$ in these samples is limited by dipole coupling between \sep\ electron spins which are not refocussed by the Hahn echo sequence (a mechanism termed instantaneous diffusion (ID)~\cite{Tyryshkin2012, Wolfowicz2012a}). This effect can be mitigated by reducing the angle of the $\theta_2$ pulse in the Hahn sequence ($\pi/2- \tau - \theta_2 - \tau - \text{echo}$)~\cite{Schweiger2001, Tyryshkin2012}, providing both a measure of the concentration of \sep electron spins, as well as a value of \ttwo\ in the absence of ID.  The concentration of \sep\ we obtain is \unit{4\times10^{13}}{cm^{-3}}, in good agreement with level of boron in this sample (noting that boron is required to ionise the selenium). The extended value of \ttwo\, in the absence of ID, can be well described by a combination of spectral diffusion arising from \tone-induced spin flips of neighbouring \sep\ spins, combined with a temperature-independent mechanism that limits $T_2$ to about \unit{80}{ms}. {\color{black} At the spin concentration used here, indirect flip-flop of neighbour electron spins is expected to limit $T_2$ to about $\sim$\unit{1}{s}~\cite{Tyryshkin2012}. We therefore postulate the temperature independent-mechanism observed arises from charge tunnelling in the sample, due to the high impurity concentration} and compensation in the material. 
Therefore, for lower donor concentrations and using alternative means of ionising the Se, we anticipate the coherence times will be at least as long as those obtained for P (and indeed, longer, for the same temperature). { \color{black} The red solid line in \Fig{T2} gives the temperature dependence of the extended value of \ttwo\ (in the absence of ID). The model is obtained by taking into account the combined effects of the three mechanisms described above: i) spin relaxation of the central spin, yielding an echo decay of the form $\exp[-(2\tau/ T_1)]$; ii) a temperature-independent process of the form $\exp[-(2 \tau / T_{2,\text{lim}})]$ with $T_{2,\text{lim}}=\unit{80}{ms}$; and iii) a spectral diffusion process caused by spin relaxation of nearest neighbours, of the form $\exp[-(2\tau/T_{SD})^2]$ with $T_{SD}^2=K\, T_1/[\text{Se}^+]$ and $K=\unit{7.6\times10^{12}}{s/cm^{3}}$\cite{Schweiger2001, Tyryshkin2012}. }

\begin{figure}[t]
\begin{center}
\includegraphics[width=250pt]{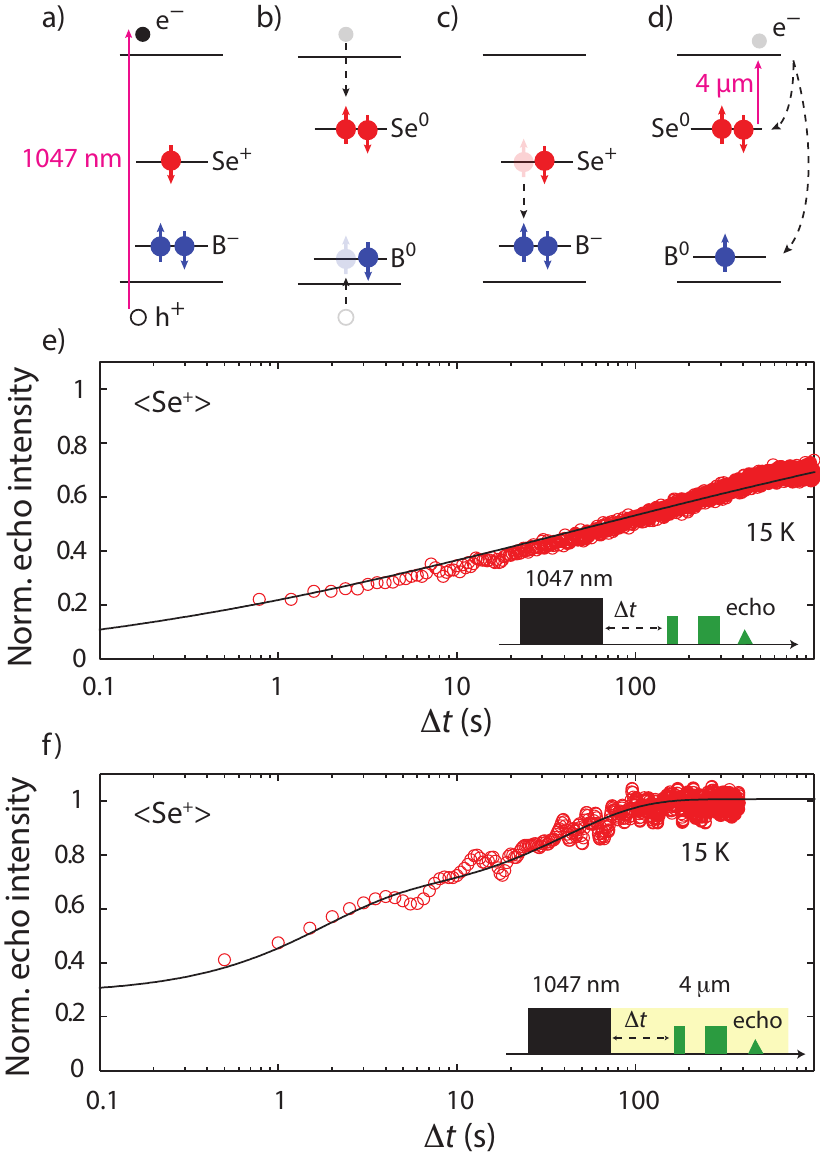}
\caption{ Schematic representation of the recombination mechanisms in selenium doped silicon after above band gap illumination: (a) At the thermal equilibrium one of the electron from a selenium impurity moves to the acceptor B forming the DA pair (Se$^+$B$^-$). Under \unit{1047}{nm} illumination electron-hole pairs are formed. (b) The excess of conduction carriers recombines through the impurities levels forming a non-equlibium state. (c) The process of DA recombination re-establish the thermal equilibrium. {\color{black} (d) Under \unit{4}{\mu m} illumination an electron from Se$^0$ is continuously pumped into the conduction band enhancing the rate at which the thermal equilibrium is re-established. The normalised concentration of selenium in the Se$^+$ state ($\langle \text{Se}^+ \rangle$) is monitored by plotting the normalised electron spin echo intensity, as a function of wait time, after a pulse of above band gap illumination, (e) in the dark, or (f) in the presence of \unit{4}{\mu m} illumination. Red circles show normalised electron spin echo intensity and black lines are fits (see main text).}\label{Relax_exp_3}}
\end{center}
\end{figure}


\section{Donor-acceptor recombination dynamics \label{recombination_exp}}

Pulsed ESR studies of donors at low temperatures often employ optical excitation to generate free carriers, as a way to `reset' the  electron spin to its thermal equilibrium state, allowing experiments to be repeated on timescales much faster than the intrinsic \tone. However, for a compensated Se-doped sample, the free electron-hole pairs generated by the optical excitation are readily captured by the \sep\ and B$^-$, and the ESR signal will only be recovered after charge equilibrium is reached through DA recombination.  

To study the dynamics of DA pairs, we measure the \sep\ electron spin echo as it recovers back to thermal equilibrium following above band-gap illumination (\unit{1047}{ nm}). This method has been previously used to study DA recombination of highly compensated phosphorus-doped silicon \cite{Dirksen1989} showing that the recombination time can be as long as $\sim\unit{10^3}{s}$.  Figure \ref{Relax_exp_3} gives a schematic representation of the DA recombination process after above band gap illumination. In this case, the system is brought out of equilibrium by the creation of free carriers via illumination (a). Due to the fast capture of such free carriers into the impurities levels, a non-equilibrium state is created after illumination (b). Thermal equilibrium is then re-established by DA recombination (c). 

We first measured the photoconductivity directly following illumination to obtain a photo-induced carrier lifetime of \unit{40}{ms} at \unit{15}{K} (process (b) in \Fig{Relax_exp_3}).  DA recombination, a much slower process, was then measured by observing the electron spin echo intensity following the laser pulse, as shown in \Fig{Relax_exp_3}(e). A pulse of \unit{100}{mW} for about 1~s was sufficient to ensure almost all the selenium and boron impurities were in the neutral state (i.e.\ zero initial spin echo signal). After $\sim$1000 seconds following the laser pulse, 60-70\% of the echo intensity had recovered indicating that the majority of selenium-boron pairs had recombined. However, obtaining the full echo intensity required a waiting time of $\sim\unit{10^4}{s}$ --- such non-exponential behaviour is consistent with a random distribution of donor-acceptor nearest-neighbour distances. 

The problem of DA recombination has been solved analytically for shallow donors by Thomas \emph{et al.}~\cite{Thomas1965} and has been adapted here to selenium in the case of low compensation ([Se]$\gg$[B]). In this case, the relative distance between donors and acceptors is much larger than the Bohr radius of both electron and hole bound the to impurities and the wavefunctions in the ground state may be considered unperturbed. For instance, for concentrations of $\sim\unit{10^{15}}{cm^{-3}}$ and low compensation, the most probable distance between DA nearest-neighbours is about \unit{50}{nm} which is much larger than $a^*_H=\unit{4.2}{nm}$, the effective Bohr radius of the bound hole in silicon. Under this approximation, the calculation of the recombination rate ($W$) for each individual DA pair only involves the optical matrix element $M$ between the wavefunctions of the donor and acceptor in the neutral condition and we have $W(r)=|M|^2$. The optical matrix element can be then taken to be proportional to the value of the hole wavefunction at the donor site, giving 
\begin{equation}
W(r)=W_{0} \exp\left\{-\frac{2r}{a^*_H}\right\}, \label{decay_rate}
\end{equation}
where $W_{0}$ has been determined to be between $\unit{10^{5}}{s^{-1}}$ and $\unit{10^{3}}{s^{-1}}$ \cite{Dirksen1989}. The effect of an ensemble of selenium impurities surrounding the acceptor can be seen as set of independent recombination channels with a certain distribution of recombination rates. By integrating over a random distribution of donors, the dynamics can be shown to follow:
\begin{equation}\label{dynamics_acceptors}
\frac {\langle \text{B}^0 (t) \rangle} {[\text{B}]} =\exp \left[4\pi [\text{Se}]   \int_0^\infty \left\{   e^{ - W(r) t }-1\right\} \,\, r^2dr \right],
\end{equation} 
\begin{equation}\label{dynamics_donor}
\frac {\langle \text{Se}^+ (t) \rangle} {[\text{B}]}   = 1- \frac {\langle \text{B}^0 (t) \rangle} {[\text{B}]} 
\end{equation}
{\color{black} where $\langle \text{Se}^+ (t)\rangle$ and $\langle \text{B}^0 (t) \rangle$ are the concentrations of Se$^+$ and B$^-$ respectively}. We simulated the dynamics of the recombination process using the equations above, taking the known concentration of selenium impurities, and find a good fit to the experimental data. The value of the relaxation rate $W_{0}$ used in the fit was $\unit{10^4}{s^{-1}}$, at \unit{15}{K}, which is consistent with previous values observed for similar samples doped with shallow donors~\cite{Dirksen1989}. 

{\color{black} The approach to thermal equilibrium (all boron ionised and an equal number of \sep) can be accelerated over that resulting from DA pair recombination by using infrared radiation at about} \unit{4}{\mu m}: an electron is continuously excited from the neutral selenium into the conduction band, whereupon it either returns to re-neutralise the selenium, or is captured by an acceptor (\Fig{Relax_exp_3}(d)). 
We find that under weak (\unit{10}{\mu W})  illumination at \unit{4}{\mu m}, {\color{black} the rate at which equilibrium charge conditions are established is two orders of magnitude faster then the direct DA recombination}. Within the first second after the \unit{1047}{nm} flash, about 50 \% of the echo signal is recovered and  thermal equilibrium state is reached after about \unit{100}{s}. The recombination is well fit by a bi-exponential decay with $\tau_1=\unit{1.6}{s}$ and $\tau_2=\unit{40}{s}$. 
The effect of \unit{4}{\mu m} illumination is further evidence that the slow observed recombination dynamics are limited by long-lived (i.e.\ distant) DA pairs. Moreover the illumination sequence ($\unit{1047}{nm}+\unit{4}{\mu m}$) could be used in order to study the {\color{black} nuclear spin properties of neutral selenium, $^{77}$Se$^0$}, following analogous methods to those used to study ionised P nuclear spins through electrically detected magnetic resonance~\cite{Dreher2012}. In shallow donors such as P, the hyperfine interaction with the electron spin strongly limits the nuclear spin coherence, such that nuclear spin coherence times for P$^+$ are strongly enhanced~\cite{Saeedi2013}. Similarly, one could expect the nuclear spin coherence time for neutral selenium (in its singlet ground state) to be comparably long.

\section{Conclusions}
{\color{black} We have found that the electron spin relaxation times of \sep\ are the longest observed for donors in 28-silicon --- this is explicitly demonstrated for temperatures in the range 5--30~K, and is expected to hold true for lower temperatures. }The energy splitting, $\Delta$, between the ground and first-excited valley state is an order of magnitude larger in \sep\ than for Group V donors in silicon --- this effectively removes the Orbach spin relaxation mechanism, though its effect in reducing the strength of the Raman $T^9$ process is rather weak,  providing motivation for further theoretical work to understand this quantitatively. {\color{black} As a result of this increase in \tone, the electron spin coherence times in \sep\ are shown to be significantly longer than for Group V donors for temperatures above 10~K, and this trend is likely to extend to lower temperatures for samples with lower Se concentrations than those studied here. 
}The large value of $\Delta$ should also result in a dramatically reduced Stark shift~\cite{Pica2014} compared to shallow donors, such that the electron and nuclear spin coherence of \sep\ could be relatively unaffected by charge noise in nano-devices, despite the large value of the hyperfine coupling.  {\color{black} It is also possible to spectrally resolve the ground state hyperfine coupling in $^{28}$Si:$^{77}$\sep\ using the $1s(T_2)\Gamma_7$ absorption transition~\cite{Steger2009}, so if a tuneable source were available at $\unit{\sim2.9}{\mu m}$ this could be used for fast and efficient hyperpolarization of both the electron and nuclear spin.}
Finally, we have investigated optical methods to manipulate the donor charge state (from Se$^+$ to Se$^0$), examining this through donor-acceptor recombination following above band gap illumination, and showing how Se$^0$ can be ionised using \unit{4}{\mu m} illumination. This suggests a possible mechanism to first bring selenium donors into an ESR-active \sep\ state, and subsequently neutralise them leaving a potentially long-lived $^{77}$Se nuclear spin.\\

\acknowledgments
We thank Joseph Smith, Pierre-Andr\'{e} Mortemousque and Cheuk Lo for useful discussions. This research is supported by the EPSRC through the Materials World Network (EP/I035536/1) and UNDEDD project (EP/K025945/1) as well as by the European Research Council under the European Community�s Seventh Framework Programme (FP7/2007- 2013)/ERC grant agreement No. 279781. Work at Princeton was supported by NSF through Materials World Network (DMR-1107606) and through the Princeton MRSEC (DMR-01420541). J.J.L.M. is supported by the Royal Society.

\bibliography{Selenium_paper.bib}

\end{document}